\documentclass[aps,reprint,onecolumn,floatfix,nofootinbib,12pt]{revtex4-2}
\usepackage{hyperref}
\hypersetup{
  breaklinks=true, % splits links across lines
  colorlinks=true, % displays links as colored text instead of blocks
  pdfusetitle=true, % \title and \author values into pdf metadata
}
\usepackage{float}
\usepackage{epsfig}
\usepackage{graphicx}
\usepackage{amsmath}
\usepackage{amsfonts}
\usepackage{amssymb}
\usepackage{bm}
\usepackage{bbold}
\usepackage{epsfig}
\usepackage{graphicx}
\usepackage{color}
\usepackage{pictex}
\usepackage{mathtools}
\usepackage{extarrows}
\usepackage{footnote}
\usepackage{footmisc}
\usepackage{dsfont}
\usepackage{setspace}
\newcommand{\s}{\scriptscriptstyle}
\begin{document}
\title{Unwithered Majorana fermions in the bulk of a quantum chain}
\author{Gennady Y. Chitov}
\affiliation{Bogoliubov Laboratory of Theoretical Physics,
Joint Institute for Nuclear Research, Dubna 141980, Russia}
\date{\today}

%
%xxxxxxxxxxxxxxxxxxxxxxxxxxxxxxxxxxxxxxxxxxxxxxxxxxxxxxxxxxxxxxxxxxxxxxxxxxxxxx
%
\begin{abstract}
The proposal is to probe Majorana modes in the system when its parameters are tuned to bring it into a disentangled ground state, which occur on the parametric curves known as disorder lines (DL). In such state certain correlation functions do not depend on separation. The exact results are presented for the XY spin chain in transverse field, also called the Kitaev chain in the Majorana representation. The single Majorana modes are shown to be localized near the ends of the chain, as in the states off the DL, while the disentangled $n$-particle Majorana modes ($n \geq 2$) penetrate into the bulk without attenuation. The predicted bulk-edge effects can be detected in the specially engineered optical chains and lattices.
\end{abstract}
\maketitle

\textit{$\bullet$ Rationale:}
The Majorana fermion \cite{Majorana:1937,*Majorana:2006} is a fascinating conjectured particle which is its own antiparticle. Its detection would bring a certain harmony in Nature: as the bosons have, e.g., photons which are their own antiparticles, the fermions would acquire their counterpart. Among the fundamental particles of the Standard Model the best (and actually the only, since it must be charge-less) candidate for the Majorana fermion is neutrino. However the experimental verification of the neutrino's Majorana nature, through, e.g., the neutrinoless double $\beta$-decay is still work in progress, no definite answer reported up to date (for a recent review, see \cite{Agostini:2023}).

Another possibility to find Majorana fermions can be explored in the context of the condensed matter physics, where they can be detected as excitations
(quasi-particles), in analogy between photons and phonons. This was clearly spelled out in the very influential paper by Kitaev \cite{Kitaev:2001} who identified the zero-energy edge states in the quantum chain as Majorana fermions. The search of such Majorana modes is a very active field of current research, but so far there are no clean-cut experimental results confirming detection of them. (For reviews and more references, see, e.g., Refs.~\cite{Alicea:2012,Lutchyn:2018,Rachel:2025,Kouwen:2025}.)

The major difficulty with the detection of the Majorana fermions (we restrict our discussion to the simplest case of one spatial dimension) is that when they presumably exist, they are localized modes (i.e., their wave-function is normalizable) only in the gapped phases. Moreover, the single-particle and n-particle ($n \geq 2$) modes are localized near the ends of the system, and decay exponentially quickly in the bulk, with the characteristic length defined by the inverse gap.

The main idea of this paper is to explore the Majorana fermions in the system when its parameters are tuned to bring it to a special disentangled state at zero temperature ($T=0$). Such ``classical'' factorized states  \cite{Muller:1982,*Muller:1985} occur on the special parametric curves, originally called disorder lines (DLs) by Stephenson \cite{Stephenson-I:1970,*Stephenson:1970PRB}. The DLs can be directly related to the behavior of zeros of the partition function (Lee-Yang zeros \cite{YangLee:1952,*LeeYang:1952}). The latter control critical properties of thermodynamic systems \cite{Huang:1987}. At the DL the system does not undergo a proper phase transition, since the Lee-Yang zeros do not become real in the thermodynamic limit. DL is a weak feature, accompanied by a cusp in correlation length, and correlation functions acquiring modulation with an incommensurate wavelength. The reason is that the couple of the Lee-Yang zeros merge on the DL, and become complex conjugate when crossing it \cite{Chitov:2017PRE,Chitov:2021}. As a consequence, some $n$-point correlation functions become constant, that is do not depend on the separation between the points. In simple terms, in the disentangled state on the DL, the Majorana modes at the edge can penetrate into the bulk without attenuation.

\textit{$\bullet$ Model and properties:}
To keep our analysis in line with the original arguments due to Kitaev \cite{Kitaev:2001}, we consider the same model. It is probably the simplest one to get the desired results. In the spin representation it is the well-known quantum $XY$ chain in transverse magnetic field \cite{LiebSM:1961,McCoyII:1971,Franchini:2017} with the Hamiltonian:
\begin{equation}
\label{XYHam}
   H = -\sum_{n=1}^{N}
   \Big\{  \frac{J}{4} \big[ (1+\gamma) \sigma_{n}^{x } \sigma_{n+1}^{x}
 + (1-\gamma) \sigma_{n}^{y} \sigma_{n+1}^{y} \big] +  \frac12 h \sigma_{n}^{z} \Big\} ~,
\end{equation}
where $\sigma^\alpha_n$ are the standard Pauli matrices, and coupling $J>0$ is ferromagnetic.
The Jordan-Wigner transformation (JWT)  maps \eqref{XYHam} onto the chain of the spinless Dirac ($c_n$) fermions (see Appendix \ref{App-1} for details), which in their turn
can be represented via Majorana fermions defined as
\begin{equation}
\label{Maj}
   a_n +i b_n  \coloneqq 2 c^{\dag}_n
\end{equation}
with the anticommutation relations
\begin{equation}
\label{MComm}
  \{a_n,a_m\} =  \{b_n,b_m\}= 2 \delta_{nm}, ~~\{a_n,b_m\} = 0
\end{equation}
In the Majorana representation  the Hamiltonian  reads (from now on we set $J=1$):
\begin{equation}
\label{XYMaj}
   H = -\frac{i}{4}  \sum_{n=1}^{N} \big[ (1+ \gamma) b_n a_{n+1} -(1 -\gamma) a_n b_{n+1}+ 2 h b_n a_n \big] ~,
\end{equation}
and $h$ plays a role of the chemical potential.

\begin{figure}[h]
\centering{\includegraphics[width=9.0cm]{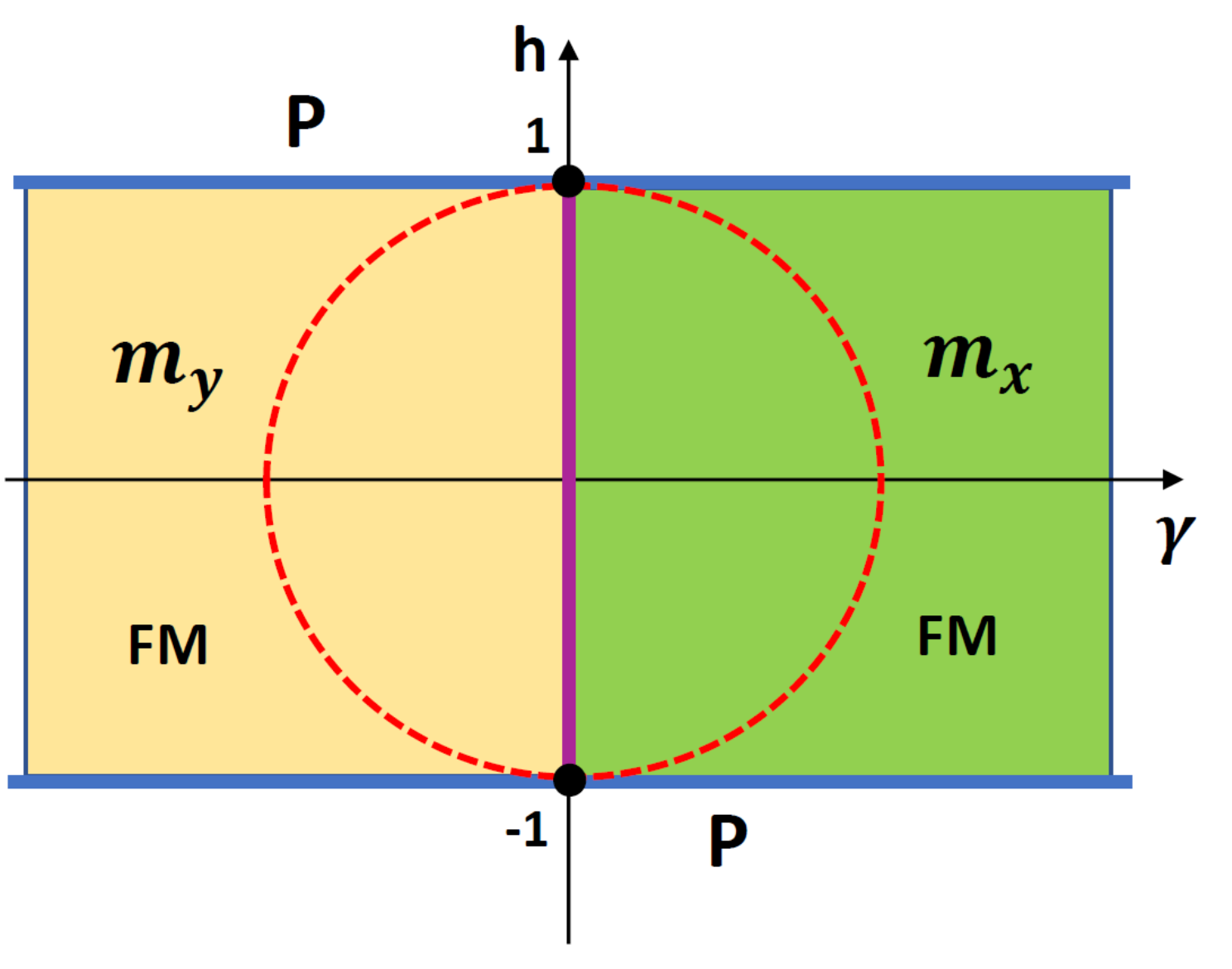}}
\caption{The phase diagram of the model on the $(\gamma,h)$ plane at $T=0$.
It has three gapped phases: polarized (P) $|h|>1$; ferromagnetic (FM) $|h|<1$; $m_x \neq 0~(\gamma>0)$ and $m_y \neq 0~(\gamma<0)$.
The model is critical (gapless) on the phase boundaries: \textit{(i)} two infinite
lines $h= \pm 1$ (dark blue); \textit{(ii)}  segment $|h| \leq 1$ at  $\gamma=0$ (magenta).
The disorder line is the circle $h^2+\gamma^2 =1$ (dashed red).}
\label{PhaseDiag}
\end{figure}

For further convenience we give explicit formulas for the spin-Majorana JWT:
\begin{eqnarray}
  \sigma_{n}^{x } \sigma_{n+1}^{x} &=& i b_n a_{n+1} \nonumber \\
  \sigma_{n}^{y} \sigma_{n+1}^{y} &=& -i a_n b_{n+1} \nonumber \\
  \sigma_{n}^{z}  &=& i b_n a_n
\label{JWTSM}
\end{eqnarray}
and
\begin{eqnarray}
\label{JWsigmaX}
  \sigma_n^x &=& (-1)^{n-1}O_z (n-1)  a_n \\
\label{JWsigmaY}
  \sigma_n^y  &=& (-1)^{n-1} O_z (n-1) b_n~.
\end{eqnarray}
We used above the $z$-string operator defined as:
\begin{equation}
\label{Oz}
  O_z(n) \coloneqq \prod_{l=1}^{n} \sigma_l^z =\prod_{l=1}^{n} \big[ i b_{l} a_{l} \big]~.
\end{equation}
In this work we consider the case $T=0$. The ground-state phase diagram of the model is shown in Fig.~\ref{PhaseDiag}. It is symmetric with respect $h \leftrightarrow -h$. More details of its  properties and exact results are provided in the Appendix \ref{App-1}.  In the spin language, at $\gamma \gtrless 0$ ($|h|<1$) the model is in the ordered gapped phase with spontaneous longitudinal magnetization $m_x/m_y$. The quantum critical line at $\gamma=0$ and $|h|<1$ corresponds to the gapless phase of the free gas of Dirac fermions with the Fermi sea controlled by $k_F = \arccos(-h)$, no localized Majorana states. For the purposes of this work we discuss the region $|h|<1$  where those states exist.  Since under the sign change $\gamma \leftrightarrow - \gamma$: $m_x \leftrightarrow m_y$, we consider the first quadrant of the $(\gamma, h)$ plane only.

%%%%%%%%%%%%%%%%%%%%%%%%%%%%%%%%%%%%%%%%%%%%%%%%%%%%%%%%%%%%%%%%%%%%%%%%%%%%%%%%%
%%%%%%%%%%%%%%%%%%%%%%%%%%%%%%%%%%%%%%%%%%%%%%%%%%%%%%%%%%%%%%%%%%%%%%%%%%%%%%%%%
%%%%%%%%%%%%%%%%%%%%%%%%%%%%%%%%%%%%%%%%%%%%%%%%%%%%%%%%%%%%%%%%%%%%%%%%%%%%%%%%%%
%%%%%%%%%%%%%%%%%%%%%%%%%%%%%%%%%%%%%%%%%%%%%%%%%%%%%%%%%%%%%%%%%%%%%%%%%%%%%%%%%%

%
%
\textit{$\bullet$ Disorder line:}
For the Majorana Hamiltonian  we construct two factorized states on the DL
\begin{equation}
\label{DLs}
  \gamma^2 +h^2 =1
\end{equation}
as two ordered vectors:
\begin{equation}
\label{GSpmMaj}
  {\left| \Psi^{\pm}  \right\rangle} = \bigotimes_{n=1}^{N} \left| \Psi^{\pm}_n  \right\rangle, ~~
   \left| \Psi^{\pm}_n  \right\rangle = \Psi^{\pm}_n \left| 0_n  \right\rangle,~~
  \Psi^{\pm}_n = a_n \cos \theta\pm \sin \theta
\end{equation}
with
\begin{eqnarray}
\label{NormLoc}
 \left\langle  \Psi^{\pm}_n  | \Psi^{\pm}_n  \right\rangle &=& 1~,~~
 \left\langle  \Psi^{-}_n   | \Psi^{+}_n  \right\rangle =\cos 2 \theta~,\\
\label{NormGlob}
 \left\langle  \Psi^{\pm}  | \Psi^{\pm}  \right\rangle &=&  1~,~~
 \left\langle  \Psi^{-}  | \Psi^{+}  \right\rangle= \cos^N 2 \theta~,
\end{eqnarray}
and
\begin{equation}
\label{Flip}
 i b_n a_n \left| \Psi^\pm _n \right\rangle = \left| \Psi^\mp _n \right\rangle
\end{equation}
The angle $\theta$ is related to the Lee-Yang roots \eqref{lampm} merging on the DL as:
\begin{equation}
\label{ThetaLam}
  \lambda _{+} =\lambda _{-} =\sqrt{\frac{1-\gamma }{1+\gamma } } =\cos 2\theta
\end{equation}
The vacuum of the Dirac fermions \eqref{Maj} is defined as
\begin{equation}
\label{Vac}
  \left| 0 \right\rangle = \bigotimes_{n=1}^{N} \left| 0_n  \right\rangle, ~~a_n  \left| 0_n  \right\rangle =i b_n  \left| 0_n  \right\rangle~.
\end{equation}
The scalar product of two vectors, defined in $2^N$-dimensional space \eqref{GSpmMaj} as
\begin{equation}
\label{ADLGS}
  \left| A \right\rangle = \bigotimes_{n=1}^{N} \left| A_n  \right\rangle, ~~
  \left| A_n  \right\rangle = \hat A_n \left| 0_n  \right\rangle
\end{equation}
is given by
\begin{equation}
\label{AB}
    \left\langle B  | A \right\rangle=
    \prod_{n=1}^N  \langle 0_n |\hat B_n \hat A_n |0_n  \rangle~,
\end{equation}
where we assumed $\hat B_n,\hat A_n$ to be self-adjoint.
Two ground states are normalized, while their overlap  $\left\langle  \Psi^{-}  | \Psi^{+}  \right\rangle$ falls off $\propto \exp(-N/\xi)$ with the correlation length
\begin{equation}
\label{CorrL}
  \xi^{-1} = -\ln \cos 2 \theta ~,
\end{equation}
and vanishes in the thermodynamic limit,  when these states become orthogonal.

To calculate the average of Maiorana operators in the ground state \eqref{GSpmMaj}, we will
act by those operators on the ket $\left| \Psi^{\pm}  \right\rangle$ and order it as in \eqref{ADLGS}. The following formulae are useful to carry out the calculation:
moving of a Majorana operator to the right
\begin{equation}
\label{Mmove}
 m \neq n:~ M_n \Psi^{\pm}_m=-\Psi^{\mp}_m M_n,~ M=a,b~;
\end{equation}
and local scalar products
\begin{eqnarray}
\label{ALocAv}
 \left\langle  \Psi^{\pm}_n   \right|  a_n \left| \Psi^{\pm}_n  \right\rangle &=& \pm \sin 2 \theta~,~~
 \left\langle  \Psi^{\pm}_n   \right|  a_n \left| \Psi^{\mp}_n  \right\rangle =0~,\\
\label{BLocAv}
 \left\langle  \Psi^{\pm}_n   \right|  b_n \left| \Psi^{\pm}_n  \right\rangle &=&  0~,~~
 \left\langle  \Psi^{-}_n   \right| i b_n \left| \Psi^{+}_n  \right\rangle= \sin 2 \theta~;
\end{eqnarray}
From all above along with the JW mapping \eqref{JWsigmaX} and \eqref{JWsigmaY} we recover results \eqref{mxDL}, \eqref{mzDL}, \eqref{myDL}, \eqref{SpinCorr}, \eqref{HPM}
obtained in the spin representation (cf. Appendix \ref{App-2}), using now the fermionic factorized states \eqref{GSpmMaj}, no calculation of the Toeplitz determinants is necessary. All other results for the Majorana operators pertinent to the further presentation are also collected in Appendix \ref{App-2}.

Using Eqs. \eqref{JWsigmaX}, \eqref{ALocAv} and \eqref{Aav} we get the following at the left end of the chain:
\begin{equation}
\label{a1}
  \bar a_1  \coloneqq  \left\langle  \Psi^{\pm}   \right|  a_1 \left| \Psi^{\pm}  \right\rangle =
  \left\langle  \Psi^{\pm}_1   \right|  a_1 \left| \Psi^{\pm}_1  \right\rangle =
  \left\langle  \Psi^{\pm}   \right|  \sigma_1^x \left| \Psi^{\pm}  \right\rangle = m_x= \pm \sin 2 \theta~,
\end{equation}
and $\bar a_n  \coloneqq \left\langle  \Psi^{\pm}   \right|  a_n \left| \Psi^{\pm}  \right\rangle= (-1)^{n-1}  \bar a_1 \exp (-(n-1)/ \xi)$.
The set $\bar a_n$  yields the coordinate representation of the wave function of $a$-Majorana fermion in the state $\left| \Psi^{+}  \right\rangle$ (or $\left| \Psi^{-}  \right\rangle$). This wave function is localized near the left end of the chain, and normalized ($\sum_n \bar a_n^2=1$)  in the limit $N \to \infty$.

Appearance of non-zero Majorana average $\bar a_n \neq 0$ signals spontaneous $\mathds{Z}_2$ symmetry breaking, and $\bar a_1$ is equal to the order parameter. Indeed, from the
spin-Majorana duality we get correspondence between the local magnetic and non-local string orders as
\begin{equation}
\label{SOPDL}
  \forall n:~~(-1)^{n-1} \left\langle \Psi^{\pm}  \right|  \prod_{l=1}^{n-1} \big[ i b_{l} a_{l} \big] a_n  \left| \Psi^{\pm}  \right\rangle =  \bar a_1~,
\end{equation}
so two vectors \eqref{GSpmMaj} correspond to the degenerate ground states with opposite signs of the order parameter.  The  ground-state energy is two-fold degenerate
\begin{equation}
\label{HPM}
 \mathcal{E}_\circ=  \left\langle \Psi^{\pm}  \right|  H  \left| \Psi^{\pm}  \right\rangle =
  -\frac{N}{4} \Big[ (1+\gamma)\bar a_1^2+ 2 h \sqrt{1-\bar a_1^2} \Big]=-\frac{N}{2}
\end{equation}
and constant on the DL.

Although the ground-state average of $b_n$ is zero in this phase, the order $\bar a_1 \neq 0$ is accompanied by appearance of $\tilde b_n  \coloneqq  \left\langle  \Psi^{-}   \right| b_n \left| \Psi^{+}  \right\rangle \neq 0$ (cf. Eqs.~\eqref{Bav}  and \eqref{Bover}). The wave function $i \tilde b_n$ (with $|\bar a_n|= | \tilde b_{N+1-n}|$) is localized near the right end, where
\begin{equation}
\label{bN}
 i \tilde b_N =  (-1)^{N-1}   \left\langle  \Psi^{-}_N   \right| i b_N \left| \Psi^{+}_N  \right\rangle
 =(-1)^{N-1}   \left\langle  \Psi^{\pm}_N   \right|  \pm a_N \left| \Psi^{\pm}_N  \right\rangle = (-1)^{N-1} \sin 2 \theta ~.
\end{equation}
We used above Eq.~\eqref{Flip}, written as
%\begin{equation}
%\label{baFlip}
 $ i b_n \left| \Psi^{\pm}_n  \right\rangle = -a_n \left| \Psi^{\mp}_n  \right\rangle$.
%\end{equation}
In the limit $N \to \infty$,  $|\tilde b_n|^2$ yields the probability to find at the node $n$ the $b$-Majorana fermion tunneling between two ground states.
There is no reason to connect these Majoraba states with a particular wave number, since they are delocalized within the Brullouin zone ($N \to \infty$):\footnote{\label{MBZ} The symmetry transformation \eqref{Pi3Tr} would bring the maximum of the distribution \eqref{ProbABq} from the edge to the center of the BZ, but  would not reduce the problem to a single ($k=0$) Majorana state.}
 \begin{equation}
\label{ProbABq}
  |  \bar a(k) |^2=|  \tilde b(k) |^2= \frac{1}{2 \pi} \frac{\sin^2 2 \theta}{1+\cos^2 2 \theta+ 2 \cos 2 \theta \cos k}~.
\end{equation}

The single $(a/b)$-Majorana states even on the DL die off exponentially with the distance from the (left/right) edge, as at a generic point on the $(h,\gamma)$-plane $(|h|<1)$, see  \cite{Karevski:2000}. To qualitatively understand why it is so, let us recall that to calculate the average as defined in \eqref{ADLGS} and \eqref{AB}, one needs to bring $a_n$ to the $n$-th position of the ket in \eqref{ADLGS}. It involves not only a trivial sign change, but also flipping of all vectors on the left of $n$: $|\Psi^{\pm}_n \rangle \mapsto  |\Psi^{\mp}_n \rangle$, cf. Eq.~\eqref{Mmove}. So, the exponential decrease of $\bar a_n$  with $n$ is due to the contributions from the overlap between $n-1$ local $\pm$ states. Similarly, $i \tilde b_n$ gets the contribution from the overlap between $\pm $ states on the right of $n$.  That is why the probability to detect a single $a$-Maiorana state is maximal at the left end,
where the overlap is absent, and so for the tunneling $b$-Majorana fermion \eqref{bN} at the right end. The same arguments explain the exponential decline of the correlation function of two Majorana fermions  \eqref{MajGFmnB}.

The correlation function of the order parameter \eqref{SOPDL}, which is the average of two Majoranas connected by a string, is constant on the DL:
\begin{equation}
\label{StrCFunDL}
  \forall m>n+1:~~(-1)^{n+m} \left\langle \Psi^{\pm}  \right| a_n  \prod_{l=n+1}^{m-1} \big[ i b_{l} a_{l} \big] a_m  \left| \Psi^{\pm}  \right\rangle =  \bar a_1^2~,
\end{equation}
as the $xx$ spin correlation functions \eqref{SpinCorr}.

The ground states $| \Psi^{\pm} \rangle$  on the DL are tensor products of the states $|\Psi^{\pm}_n \rangle$ which are ``locally ordered''  with the order parameter $\bar a_1=\pm \sin 2 \theta$, cf. Eq.~\eqref{ALocAv}. The model is disentangled on the DL, since the concurrence is shown to be zero \cite{Chitov:2021}.
The state vector $|\Psi^{\pm}_n \rangle$ at each node $n$ is a superposition of empty and occupied states of the Dirac fermion: $\left| \Psi^{\pm}_n  \right\rangle = |1_n\rangle \cos \theta \pm  |0_n\rangle \sin \theta$. That is: no definite parity, irrational occupation number $\cos^2 \theta$.
The formulae are applicable at any finite $N$, as well as in the thermodynamic limit $N \to \infty$, no need to wonder whether the Avogardo number is even or odd.

The Ising point on the DL ($\gamma=1$, $h=0$, $\theta=\pi/4$) corresponds to the maximal value of the order parameter $|\bar a_1|=1$. $|\Psi^{\pm}_n \rangle$  at each (half-filled) site is an equally weighted superposition of $|1_n\rangle$ and  $|0_n\rangle$. The other end of the DL ($\gamma=0$, $h=1$, $\theta=0$) merges with the critical point of the P-phase. The order parameter $\bar a_1=m_x \to 0$, the states $|\Psi^{+}_n \rangle \to |\Psi^{-}_n \rangle \to |1_n\rangle$ merge to a singly occupied state (eigenstate of parity +1).  Such unique disentangled fully polarized ground state $\otimes_n |1_n\rangle $ exists on the whole line  $\gamma=0$, $h \geq 1$.

\textit{$\bullet$ Observables (Dirac fermions):}
For the experimental verifications, it is important that in the ground states on the DL, the Dirac fermion is effectively split into a single $a$ or $b$ Majorana fermion:
\begin{eqnarray}
\label{ccdagAv}
   \left\langle  \Psi^{\pm}   \right|  c_n \left| \Psi^{\pm} \right\rangle  &=& \left\langle  \Psi^{\pm}   \right|  c_n^\dag \left| \Psi^{\pm} \right\rangle  =
  \frac12 \left\langle  \Psi^{\pm}   \right|  a_n \left| \Psi^{\pm} \right\rangle~, \\
\label{ccdagOver}
   \left\langle  \Psi^{-}   \right|  c_n^\dag \left| \Psi^{+} \right\rangle  &=& -\left\langle \Psi^{-}   \right|  c_n \left| \Psi^{+} \right\rangle  =
  \frac12 \left\langle  \Psi^{-}   \right|  i b_n \left| \Psi^{+} \right\rangle~,
\end{eqnarray}
along with ($m>n$):
\begin{eqnarray}
\label{2DAv}
   \left\langle  \Psi^{\pm}   \right|  c_n^\dag c_m \left| \Psi^{\pm} \right\rangle  &=& \left\langle  \Psi^{\pm}   \right|  c_n^\dag c_m^\dag \left| \Psi^{\pm} \right\rangle  =
  \frac14 \left\langle  \Psi^{\pm}   \right| i b_n a_m \left| \Psi^{\pm} \right\rangle~, \\
\label{2DOver}
   \left\langle  \Psi^{-}   \right|  c_n^\dag c_m^\dag \left| \Psi^{+} \right\rangle  &=& -\left\langle \Psi^{-}   \right| c_n^\dag c_m \left| \Psi^{+} \right\rangle  =
  \frac14 \left\langle  \Psi^{-}   \right|  i  a_n b_m \left| \Psi^{+} \right\rangle~,
\end{eqnarray}
Thus the averages of one and two Dirac fermions on the DL probe the Majorana states. However these parameters are localized near the ends of the chain.

To get a full advantage of being in the disentangled state on the DL, one needs to probe an average of more than two Majorana operators.
The string parameters  \eqref{SOPDL}  and \eqref{StrCFunDL} stay constant from the edge to the bulk, but their measurement seem to be a very challenging experimental problem in practice \cite{Bloch:2017}. Possibly the simplest non-trivial task would be to probe the constant $4$-point Majorana correlation function:
\begin{eqnarray}
\label{4MCorr}
\forall m>n+1:~ \mathfrak{G}_{nm} &=& \left\langle  \Psi^{\pm}   \right| i b_n a_{n+1}  i b_m a_{m+1} \left| \Psi^{\pm}  \right\rangle \nonumber \\
&=& \left\langle  \Psi^{\pm}   \right| i b_n a_{n+1}  \left| \Psi^{\pm}  \right\rangle
\left\langle  \Psi^{\pm}   \right|   i b_m a_{m+1} \left| \Psi^{\pm}  \right\rangle
= \sin^4 2 \theta = \bar a_1^4~.
\end{eqnarray}
$\mathfrak{G}_{nm}$ is related to the 4-point (normal and anomalous) correlation functions  of Dirac fermions as follows  ($m>n+1$):
\begin{eqnarray}
\label{4DCorr}
 \left\langle  \Psi^{\pm}   \right| c_n^\dag c_{n+1}  c_m^\dag c_{m+1} \left| \Psi^{\pm}  \right\rangle &=&
 \left\langle  \Psi^{\pm}   \right| c_n^\dag c_{n+1}  \left| \Psi^{\pm}  \right\rangle
  \left\langle  \Psi^{\pm}   \right| c_m^\dag c_{m+1} \left| \Psi^{\pm}  \right\rangle = \frac14 \mathfrak{G}_{nm}~, \nonumber \\
\left\langle  \Psi^{\pm}   \right| c_n^\dag c_{n+1}^\dag  c_m^\dag c_{m+1}^\dag \left| \Psi^{\pm}  \right\rangle &=&
 \left\langle  \Psi^{\pm}   \right| c_n^\dag c_{n+1}^\dag  \left| \Psi^{\pm}  \right\rangle
  \left\langle  \Psi^{\pm}   \right| c_m^\dag c_{m+1}^\dag \left| \Psi^{\pm}  \right\rangle = \frac14 \mathfrak{G}_{nm}~.
\end{eqnarray}
In the spin language $\mathfrak{G}_{nm}$ is known as the dimer-dimer (four-spin) correlation function.

The predicted effects can be detected in the specially engineered optical chains and lattices with controllable parameters of the microscopic spin or fermionic (Dirac) Hamiltonians \cite{Bloch:2008,*Bloch:2016,*Bloch:2025}.

\textit{$\bullet$ Concluding remarks:}
The ground-state averaging of a single $b_n$-Majorana operator always yields zero, it appears only as an overlap between the ground states  $\left| \Psi^{\pm}  \right\rangle$. So, $b$-Majorana can be identified as a kink zero mode \cite{Rajaraman:1987}. However it contributes to the results for the observables on the same footing with the $a$-Majorana. Although the final results are expressed via powers of the order parameter $\bar a_1$, they cannot be obtained as an average of $n$ $a$-operators only, because such average always vanishes, cf. Eqs.~\eqref{AABB}. We attribute this to the non-local nature of the Majorana order.
On the other hand, the averaged ground state energy  $\mathcal{E}_\circ (\bar a_1)$ is Landau-like, regardless: the order parameter \eqref{a1} can be obtained from minimization of \eqref{HPM}.

In case of $\gamma<0$, all the results hold with the exchange $a_n \leftrightarrow i b_n$.

%
%xxxxxxxxxxxxxxxxxxxxxxxxxxxxxxxxxxxxxxxxxxxxxxxxxxxxxxxxxxxxxxxxxxxxxxxxxxxxxx
%\begin{acknowledgments}
%The author acknowledges .....
%\end{acknowledgments}
%xxxxxxxxxxxxxxxxxxxxxxxxxxxxxxxxxxxxxxxxxxxxxxxxxxxxxxxxxxxxxxxxxxxxxxxxxxxxxx
%

%xxxxxxxxxxxxxxxxxxxxxxxxxxxxxxxxxxxxxxxxxxxxxxxxxxxxxxxxxxxxxxxxxxxxxxxxxxxxxx
%\section*{Data Availability}
%The additional data that support the findings of this article are available
%from the author upon reasonable request.
%xxxxxxxxxxxxxxxxxxxxxxxxxxxxxxxxxxxxxxxxxxxxxxxxxxxxxxxxxxxxxxxxxxxxxxxxxxxxxx

\newpage
%xxxxxxxxxxxxxxxxxxxxxxxxxxxxxxxxxxxxxxxxxxxxxxxxxxxxxxxxxxxxxxxxxxxxxxxxxxxxxx

\section*{Supplementary Materials}\label{App}
%%%%%%%%%%%%%%%%%%%%%%%%%%%%%%%%%%%%%%%%%%%%%%%%%%%%%%%%%%%%%%%%%%%%%%%%%%%%%%
\begin{appendix}
\section{General properties of the model}\label{App-1}
%%%%%%%%%%%%%%%%%%%%%%%%%%%%%%%%%%%%%%%%%%%%%%%%%%%%%%%%%%%%%%%%%%%%%%%%%%%%%%
%
%
The Jordan-Wigner transformation  (JWT) \cite{LiebSM:1961,DuttaTIM:2015} maps \eqref{XYHam} onto the spinless (Dirac) fermions
\begin{equation}
\label{XYFermi}
   H = - \sum_{n=1}^{N}  \Big\{  \frac{J}{2}  \big[ t_n (c_{n}^\dag c_{n+1} +\mathrm{h.c.})+ \gamma (c_{n}^\dag c_{n+1}^\dag +\mathrm{h.c.}) \big]
             + h_n \big( c_{n}^\dag c_{n}- \frac12~ \big)  \Big\} ~.
\end{equation}
The Hamiltonian \eqref{XYHam} is invariant under $(x,y)$-inversion $\hat \pi_2=\hat \pi_2^\dag=\hat \pi_2^{-1}= O_z(N)$:
\begin{eqnarray}
\label{Pi2Tr}
  \hat \pi_2 \sigma_{n}^{\alpha }   \hat \pi_2 &=& - \sigma_{n}^{\alpha }~, ~\alpha=x,y~, \nonumber \\
  \hat \pi_2 \sigma_{n}^z   \hat \pi_2 &=& \sigma_{n}^z~,
\end{eqnarray}
so
\begin{equation}
\label{Pi2Comm}
  [\hat \pi_2 , H]=0~.
\end{equation}
The Hamiltonian is also invariant under the combined parity or proper inversion transformation, defined by a unitary Hermitian operator $\hat \pi_3$ \cite{Sakurai:2013},
and the sign change of the magnetic field:
\begin{eqnarray}
\label{Pi3Tr}
  \hat \pi_3 \sigma_{n}^{\alpha }   \hat \pi_3 &=& - \sigma_{n}^{\alpha }~, ~\alpha=x,y,z~, \nonumber \\
  \hat \pi_3  H(-h_n) \hat \pi_3 &=&  H(h_n)~.
\end{eqnarray}
When $\gamma \neq 0$ the total number of fermions $\hat N_F= \sum_{n=1}^{N} c_n^\dag c_n$ (or $z$-component of the total spin) is not a conserving quantum number, since $[\hat N_F,H] \neq 0$, but $O_z(N)=(-1)^{\hat N_F-N}$ is, according to \eqref{Pi2Comm}.

The spin Hamiltonian \eqref{XYHam} is the well-known XY model in transverse field \cite{LiebSM:1961,McCoyII:1971,Franchini:2017}.
In the range $|h|<1$ it is ordered: $\langle\sigma_{L}^{x}\sigma_{L+n}^{x}\rangle \rightarrow  m_{x}^{2}$ as $n \rightarrow \infty$, with the spontaneous longitudinal
magnetization \cite{McCoyII:1971}
\begin{equation}
\label{mx}
   m_{x}^{2}=\frac{2}{1+\gamma} \big[ \gamma^2 (1-h^2 ) \big]^{1/4}
\end{equation}
being the order parameter. At $\gamma <0$ the order changes: $m_{x} \leftrightarrow m_{y}$. In the Majorana representation \eqref{XYMaj} the local magnetic order $m_{x,y}$ becomes non-local, with the string order parameters made out of strings of the Majorana fermions, cf.  Eqs.~\eqref{JWsigmaX} and \eqref{JWsigmaY}.

In the polarized (P) phase $|h|>1$: $m_x=m_y=0$, and only the induced magnetization $m_z \neq 0$. It it an odd function of magnetic field $m_z(-h)=-m_z(h)$ and evolves continuously across critical points $h= \pm1$  as
\begin{equation}
\label{mzUni}
 m_z= 2 \bar n-1=\langle ib_n a_n \rangle = \frac{1}{2 \pi}    \int_{-\pi}^{\pi}  dk \frac{h+\cos k}{\sqrt{(h+\cos k)^2+\gamma^2 \sin^2 k}}~.
\end{equation}
Here $\bar n=\langle \hat N_F\rangle/N$ is the average fermion filling per site. The order parameter for the P-phase is proposed  \cite{Chitov:2019,Chitov:2020} as a non-vanishing limit of the string-string  correlator
\begin{equation}
\mathfrak{D}_{zz}(L,R) = \Big\langle\prod_{n=L}^{R} \sigma_{n}^{z}\Big\rangle
\xrightarrow[\s R-L \to \infty]{~}  [ \mathrm{sign}(h)]^{R-L+1}  \mathcal{O}_z^2~.
\label{SOPz}
\end{equation}
The $z$-string order parameter $\mathcal{O}_z$ is found analytically \cite{Chitov:2022}:
\begin{equation}
\label{OzAn}
|h| > 1:~  \mathcal{O}_z^2= \Big[ \frac{h^2-1}{h^2+\gamma^2-1} \Big]^{1/4}~,
\end{equation}
and $\mathcal{O}_z=0$ when $|h|<1$. It vanishes at the boundary of the polarized phase with the 2D-Ising critical index:  $\mathcal{O}_z \propto (h-1)^{1/8}$. In the isotropic limit $\gamma=0$ it becomes a plateau  with a discontinuity at the phase boundary:
\begin{equation}
\label{OzPlateau}
 \gamma=0:~ \mathcal{O}_z=
  \left\{
  \begin{array}{lr}
  1,~ &|h| > 1~, \\ [0.2cm]
  0,~ &|h| < 1~, \\
  \end{array}
  \right.
\end{equation}
similar to the plateau of induced magnetization $m_z$.

The superconductivity parameter defined as
\begin{equation}
\label{Osc}
\mathcal{O}_{sc}^2 \coloneqq \langle c_{n}^\dag c_{n+1}^\dag +\mathrm{h.c.} \rangle = \frac{i}{2} \langle b_n a_{n+1} + a_n b_{n+1} \rangle
\end{equation}
reads explicitly as
\begin{equation}
\label{OscUni}
\mathcal{O}_{sc}^2 = \frac{\gamma}{ \pi} \int_0^\pi dk  \frac{ \sin^2 k}{\sqrt{(h+\cos k)^2+\gamma^2 \sin^2 k}}~.
\end{equation}
$\mathcal{O}_{sc}$ is not an order parameter to deal with the gapped phases of the model. It is continuous across the phase boundaries $P \leftrightarrow FM$, and does not vanish even at the critical points between those phases, where the gap closes. The only distinctive feature  of the superconductivity in those phases is that in addition to $\mathcal{O}_{sc}$, a localized edge Majorana mode is present in $m_{x,y}$-phases, and it disappears in $P$-phase \cite{Karevski:2000}.

We define the Fourier transforms of the fermion operators as:
\begin{equation}
\label{MajFour}
 2 c^{\dag}_n = a_n +i b_n  =\frac{1}{\sqrt{N}} \sum_{k \in BZ} e^{-ikn} [2 c^\dag(k)=a(k) +i b(k) ]~.
\end{equation}
Since $a_n,b_n$ are self-adjoint,
\begin{equation}
\label{MConj}
a^\dag (k)=a(-k),~~ b^\dag (k)=b(-k).
\end{equation}
The anticommutation relations are
\begin{equation}
\label{MCommFour}
  \{a(k),a(q)\} =  \{b(k),b(q)\} = 2 \delta_{k,-q}, ~~ \{a(k),b(q)\}= 0~.
\end{equation}

In terms of the Fourier-transformed Majoranas, the Hamiltonian \eqref{XYMaj} reads
\begin{eqnarray}
\label{MajF}
   H &=& -\frac{i}{2}  \sum_{k } D(k) b(k) a(-k),\\
\label{Dk}
   D(k) &=& h+\cos k +i \gamma \sin k =|D(k)| e^{i \vartheta_k}~,
\end{eqnarray}
with
\begin{eqnarray}
\label{ModDk}
   |D(k)| &=& \sqrt{(h+\cos k)^2+\gamma^2 \sin^2 k}
   ~,\\
   \sin \vartheta_k &=& \frac{\gamma \sin k}{\sqrt{(h+\cos k)^2+\gamma^2 \sin^2 k}}~,
\label{ThetaK}
\end{eqnarray}
and the wave numbers are restricted to the Brillouin zone $k \in [-\pi, \pi]$.

To diagonalize the Hamiltonian in the Majorana representation, we use the operator identity:
\begin{eqnarray}
\label{DiagMaj}
  \forall ~E :~ E (k)= E^\ast (k)=E (-k) \Longrightarrow \nonumber\\
  \frac{i}{2}  \sum_{k} E (k) b (k) a (-k) = \sum_{k} E (k) \Big[ c^\dag (k) c(k)-\frac12 \Big]~,
\end{eqnarray}
where  the Dirac and Majorana fermions are related as on the r.h.s. of \eqref{MajFour}.
A Bogoliubov transformation
\begin{equation}
\label{BTab}
  a(k)= \tilde a(k)~,~~ b(k)= e^{-i \vartheta_B(k)} \tilde b(k)
\end{equation}
preserves \eqref{MConj} and \eqref{MCommFour} for any odd real $\vartheta_B(-k)=-\vartheta_B(k)$, and the choice  $\vartheta_B(k)=\vartheta_k$ diagonalizes the
Hamiltonian in terms of the Bogoliubov fermions
\begin{equation}
\label{HBog}
  H=-\sum_{k} |D(k)|\Big[ \eta^\dag (k) \eta(k)-\frac12 \Big]~,~~2 \eta^\dag (k)\coloneqq \tilde a(k) +i \tilde b(k)~.
\end{equation}
Averaging over the ground state of the Bogoliubov fermions yields:
\begin{eqnarray}
\label{abk}
  \langle i b (k) a (q) \rangle &=& [2 \Theta(-E(k))-1]\delta_{k,-q}  \\
\label{aabbk}
  \langle  a (k) a (q) \rangle &=&    \langle  b (k) b (q) \rangle = \delta_{k,-q}
\end{eqnarray}
From Eqs.~\eqref{BTab} and \eqref{abk} we obtain the Majorana Green's functions:
\begin{equation}
\label{Gk}
  G(k)= \langle i b (k) a(-k) \rangle =e^{-i \vartheta_B(k)}= \sqrt{ \frac{h+\cos k -i \gamma \sin k}{h+\cos k +i \gamma \sin k}}
\end{equation}
and
\begin{equation}
\label{GnF}
   G_{n}= \langle i b_l a_{l+n} \rangle =  \frac{1}{2 \pi}    \int_{-\pi}^{\pi}  dk e^{ikn} G(k)~.
\end{equation}

Upon analytical continuation of $z \coloneqq e^{ik}$ onto a complex plane, we get:
\begin{equation}
\label{Gn}
  G_{n} = \oint _{\left|z\right|=1}\frac{dz}{2\pi i}  z^{n-1} G \left(z\right)~,
\end{equation}
where
\begin{equation}
\label{GzXY}
   G(z)
   = \left[ \frac{(1+z \lambda_+)(1+z \lambda_-)}
   {(z+\lambda_+)(z+\lambda_-)} \right]^{1/2}
\end{equation}
and
\begin{equation}
\label{lampm}
\lambda_\pm = \frac{h \pm \sqrt{h^2+\gamma^2-1}}{1+\gamma}
\end{equation}
are the Lee-Yang roots. $G(z)$ is usually called the generating function (symbol) in mathematical literature \cite{McCoy:2010}.
It has the following properties:
\begin{equation}
\label{GzInv}
   G(z)=1/G(z^{-1})~,
\end{equation}
and in addition, the symmetry transformation \eqref{Pi3Tr}
\begin{eqnarray}
\label{InvTr}
  h &\mapsto& -h:~~ \lambda_\pm \mapsto -\lambda_\mp   \nonumber  \\
  k &\mapsto& -k:~~ z \mapsto 1/z
\end{eqnarray}
leads to
\begin{equation}
\label{D+D-}
 G(z) \mapsto \tilde G(z)= \left[ \frac{(z-\lambda_+)(z-\lambda_-)}{(1-z \lambda_+)(1-z \lambda_-)}  \right]^{1/2}~,~
  G_n =(-1)^{n-1} \tilde G_{-n}~.
\end{equation}
The function $\tilde G(z)$, which was used in earlier related work, e.g., \cite{Chitov:2021}, is more convenient for calculations.

%%%%%%%%%%%%%%%%%%%%%%%%%%%%%%%%%%%%%%%%%%%%%%%%%%%%%%%%%%%%%%%%%%%%%%%%%%%%%%
\section{Model on the disorder line}\label{App-2}
%%%%%%%%%%%%%%%%%%%%%%%%%%%%%%%%%%%%%%%%%%%%%%%%%%%%%%%%%%%%%%%%%%%%%%%%%%%%%%
%
%
In the $XY$ chain a factorizable ferromagnetic state occurs on the DL $\gamma^2 +h^2 =1$. In the spin representation it can be written as \cite{Franchini:2017}:
\begin{equation}
\label{GSpm}
  {\left| \Psi^{\pm}  \right\rangle} = \bigotimes_{n=1}^{N} \left| \Psi^{\pm}_n  \right\rangle, ~~
   \left| \Psi^{\pm}_n  \right\rangle = \cos \theta~ {\left| \uparrow _{n}  \right\rangle} \pm \sin \theta~ {\left| \downarrow _{n}  \right\rangle}~,
\end{equation}
with the properties  \eqref{NormLoc}, \eqref{NormGlob}, and  \eqref{ThetaLam}.
The vectors \eqref{GSpm} correspond to the two degenerate ordered states with opposite signs of the spontaneous magnetization ($\forall n$)
\begin{equation}
\label{mxDL}
  m_x= \left\langle \Psi^{\pm}  \right|  \sigma _n^x  \left| \Psi^{\pm}  \right\rangle
  =\pm \sin 2\theta~,
\end{equation}
and
\begin{eqnarray}
\label{mzDL}
  m_z  &=& \left\langle \Psi^{\pm}  \right|  \sigma _n^z \left| \Psi^{\pm}  \right\rangle
  = \cos 2\theta~,\\
\label{myDL}
   m_y &=& \left\langle \Psi^{\pm}  \right|  \sigma_n^y  \left| \Psi^{\pm}  \right\rangle
   =0~.
\end{eqnarray}
The spin-spin correlation functions do not depend on the node  separation \cite{McCoyII:1971}:
\begin{equation}
\label{SpinCorr}
  {\left\langle \Psi^\pm  \right|} \sigma _{m}^{\alpha} \sigma _{n}^{\alpha} {\left| \Psi^\pm  \right\rangle} =m_\alpha^2~,
   ~~\forall~ m \neq n.
\end{equation}
Vectors \eqref{GSpm} yield the two-fold degenerate ground-state energy as
\begin{equation}
\label{HPMspin}
 \mathcal{E}_\circ=  \left\langle \Psi^{\pm}  \right|  H  \left| \Psi^{\pm}  \right\rangle = -\frac{N}{4}\big[ (1+\gamma) m_x^2 +2 h m_z \big]= -\frac{N}{2}~.
\end{equation}
Compare to \eqref{HPM}. From \eqref{Flip} we infer
\begin{equation}
\label{OzPM}
  O_z(N) \left| \Psi_{\pm} \right\rangle  = \left| \Psi_{\mp} \right\rangle ~,
\end{equation}
so, the string correlation function
\begin{equation}
\label{GSProd}
  \mathfrak{D}_{zz}(1,N) =\left\langle \Psi^{+} | \Psi^{-} \right\rangle
\end{equation}
measures the overlap between two ground states with opposite  spontaneous magnetization $\pm m_x$. It falls off $\propto \exp(-N/\xi)$ with the correlation length
\eqref{CorrL}, and vanishes in the thermodynamic limit, when these states become orthogonal.

To calculate the average of Maiorana operators in the ground states \eqref{GSpmMaj} on the DL, we use
Eqs.~\eqref{ADLGS}, \eqref{AB}, \eqref{Mmove}, \eqref{ALocAv}, and \eqref{BLocAv} to find:
\begin{eqnarray}
\label{Aav}
 \left\langle  \Psi^{\pm}   \right|  a_n \left| \Psi^{\pm}  \right\rangle &=& \pm (-1)^{n-1} \sin 2 \theta  \cos^{n-1} 2 \theta ~,\\
\label{Bav}
 \left\langle  \Psi^{\pm}   \right|  b_n \left| \Psi^{\pm}  \right\rangle &=&  0~,\\
\label{ABnav}
 \left\langle  \Psi^{\pm}   \right|  i b_n a_n \left| \Psi^{\pm}  \right\rangle &=&  \cos 2 \theta~,
\end{eqnarray}
and
\begin{eqnarray}
\label{Aover}
 \left\langle  \Psi^{\pm}   \right|  a_n \left| \Psi^{\mp}  \right\rangle &=&  0 ~,\\
\label{Bover}
 \left\langle  \Psi^{-}   \right| i b_n \left| \Psi^{+}  \right\rangle  &=& (-1)^{n-1}  \sin 2 \theta  \cos^{N-n} 2 \theta ~.
\end{eqnarray}
For $m>n$:
\begin{eqnarray}
\label{MajGFmnB}
 \left\langle  \Psi^{\pm}   \right| i b_n a_m \left| \Psi^{\pm}  \right\rangle &=& (-1)^{m-n-1} \sin^2 2 \theta \cos^{m-n-1} 2 \theta ~,\\
\label{MajGFmnL}
 \left\langle  \Psi^{\pm}   \right|  i a_n b_m \left| \Psi^{\pm}  \right\rangle &=&  0~,
\end{eqnarray}
and
\begin{eqnarray}
\label{ABoverB}
 \left\langle  \Psi^{-}   \right| i b_n a_m \left| \Psi^{+}  \right\rangle  &=&0~,\\
\label{ABoverL}
 \left\langle  \Psi^{-}   \right|  i a_n b_m \left| \Psi^{+}  \right\rangle  &=& (-1)^{m-n-1} \sin^2 2 \theta \cos^{N-m+n-1} 2 \theta ~.
\end{eqnarray}
For $m \neq n$:
\begin{eqnarray}
\label{AABB}
 \left\langle  \Psi^{\pm}   \right| a_n a_m \left| \Psi^{\pm}  \right\rangle=
             \left\langle  \Psi^{\pm}   \right| b_n b_m \left| \Psi^{\pm}  \right\rangle &=& 0~,\\
\label{AABBover}
 \left\langle  \Psi^{\pm}   \right| a_n a_m \left| \Psi^{\mp}  \right\rangle=
             \left\langle  \Psi^{\pm}   \right| b_n b_m \left| \Psi^{\mp}  \right\rangle &=& 0~.
\end{eqnarray}
As a cross-check: the results \eqref{MajGFmnB} and \eqref{MajGFmnL} coincide with a direct calculation of $G_n$ via \eqref{Gn} with the generating function \eqref{DzDL} using the residue theorem. As a consequence of the Lee-Yang roots \eqref{lampm} merging on the DL , the generating function \eqref{GzXY} becomes meromorphic:
\begin{equation}
\label{DzDL}
  G(z)=\frac{1+z \cos 2 \theta}{z + \cos 2 \theta}~.
\end{equation}

Contrary to the exponentially decreasing Majorana correlation function $G_{m-n}$, the average of  $i a_n b_{m}$ between different ground states \eqref{ABoverL} is growing with the separation $(m-n)$, reaching its maximum when two Majoranas are at the opposite ends of the chain:
\begin{equation}
\label{EdgeOver}
 \left\langle  \Psi^{-}   \right|  i a_1 b_N \left| \Psi^{+}  \right\rangle  = (-1)^{N} \sin^2 2 \theta  ~.
\end{equation}
The above result is consistent with \eqref{a1} and \eqref{bN}

The order parameter of the superconductivity \eqref{Osc}, found on the DL as:
\begin{equation}
\label{OscDL}
\mathcal{O}_{sc}^2 =\frac12 \left\langle  \Psi^{\pm}   \right| i b_n a_{n+1} \left| \Psi^{\pm}  \right\rangle = \frac12 \sin^2 2 \theta = \frac12 \bar a_1^2~.
\end{equation}
is in agreement with \eqref{OscUni}.

%
%xxxxxxxxxxxxxxxxxxxxxxxxxxxxxxxxxxxxxxxxxxxxxxxxxxxxxxxxxxxxxxxxxxxxxxxxxxxxxx
\end{appendix}
%xxxxxxxxxxxxxxxxxxxxxxxxxxxxxxxxxxxxxxxxxxxxxxxxxxxxxxxxxxxxxxxxxxxxxxxxxxxxxx
%xxxxxxxxxxxxxxxxxxxxxxxxxxxxxxxxxxxxxxxxxxxxxxxxxxxxxxxxxxxxxxxxxxxxxxxxxxxxxx

%%%%%%%%%%%%%%%%%%%%%%%%%%%%%%%%%%%%%%%%%%%%%%%%%%%%%%%%%%%%%%%%%%%%%%%%%%%%%%
%%%%%%%%%%%%%%%%%%%%%%%%%%%%%%%%%%%%%%%%%%%%%%%%%%%%%%%%%%%%%%%%%%%%%%%%%%%%%%
\bibliography{C:/Users/gchitov/Documents/Papers/BibRef/CondMattRefs}
%\bibliography{C:/Papers/BibRef/CondMattRefs}
%\bibliography{CondMattRefs}
%\bibliographystyle{apsrev4-1} \bibliography{ref2}
%%%%%%%%%%%%%%%%%%%%%%%%%%%%%%%%%%%%%%%%%%%%%%%%%%%%%%%%%%%%%%%%%%%%%%%%%%%%%%
%%%%%%%%%%%%%%%%%%%%%%%%%%%%%%%%%%%%%%%%%%%%%%%%%%%%%%%%%%%%%%%%%%%%%%%%%%%%%%
%
%
%\bibliography{CondMattRefs}
%\bibliographystyle{apsrev4-1} \bibliography{ref2}
%xxxxxxxxxxxxxxxxxxxxxxxxxxxxxxxxxxxxxxxxxxxxxxxxxxxxxxxxxxxxxxxxxxxxxxxxxxxxxx
%
\end{document}